\documentclass[twocolumn]{aastex63}

\received{June 1, 2019}
\revised{January 10, 2019}
\accepted{\today}
\submitjournal{ApJ}

\shorttitle{HL Tau polarization}
\shortauthors{Mori and Kataoka}
\graphicspath{{./}{figures/}}

\begin{document}

\title{Modeling of the ALMA HL Tau Polarization by Mixture of Grain Alignment and Self-scattering \footnote{Released on June, 10th, 2019}}

\author[0000-0002-0786-7307]{Tomohiro Mori}
\affil{The Institute of Astronomy, the University of Tokyo, 2-21-1 Osawa, Mitaka, Tokyo 181-8588, Japan}

\author[0000-0003-4562-4119]{Akimasa Kataoka}
\affiliation{National Astronomical Observatory of Japan, 2-21-1 Osawa, Mitaka, Tokyo 181-8588, Japan}

\nocollaboration{2}



\begin{abstract}
Dust polarization at (sub)millimeter wavelengths has been observed for many protoplanetary disks.
Theoretically, multiple origins potentially contribute to the polarized emission but it is still uncertain what mechanism is dominant in disk millimeter polarization. 
To quantitatively address the origin, we perform radiative transfer calculations of the mixture of alignment and self-scattering induced polarization to reproduce the 3.1 mm polarization of the HL Tau disk, which shows azimuthal pattern in polarization vectors. 
We find that a mixture of the grain alignment and self-scattering is essential to reproduce the HL Tau 3.1 mm polarization properties. 
Our model shows that the polarization of the HL Tau at 3.1 mm can be decomposed to be the combination of the self-scattering parallel to the minor axis and the alignment-induced polarization parallel to the major axis, with the orders of $\sim 0.5\%$ fraction for each component. This slightly eases the tight constraints on the grain size of $\sim 70~\micron$ to be  $\sim 130~\micron$ in the previous studies but further modeling is needed. In addition, the grain alignment model requires effectively prolate grains but the physics to reproduce it in protoplanetary disks is still a mystery.
\end{abstract}



\section{Introduction} \label{sec:intro}
The presence of dust polarization of protoplanetary disks at submillimeter/millimeter wavelengths have been indicated for several protostars and T Tauri stars with the James Clerk Maxwell Telescope (JCMT) single-channel polarimeter \citep{akeson1997, tamura1995, tamura1999}.
However, for two decades, spatially resolved polarization has not been obtained due to the lack of the sensitivity and spatial resolutions of the instruments \citep{Hughes2009, Hughes2013} until the first detection of the millimeter polarization for the HL Tau disk by \citep{stephens2014}. Nowadays, ALMA dust polarization observations have enabled us to obtain the polarization for various class I\hspace{-.1em}I protoplanetary disks, showing the various polarization morphology and polarization fraction \citep{kataoka2016b, kataoka2017, stephens2017, hull2018, bacciotti2018, ohashi2018, dent2019, harrison2019, mori2019}. 

One difficulty in interpreting the disk polarization is that multiple mechanisms of polarization are at work at millimeter wavelengths. One mechanism is the grain alignment, where elongated dust grains are aligned to some directions and emit polarized thermal dust emission. 
The grain alignment mechanism has been long and well studied, especially the radiative torques (RATs) \citep[e.g.,][]{lazarian2007}. Here we only discuss which direction the grains are aligned to because it reflects the polarization orientation. The directions of the grain alignment may represent magnetic fields \citep{davis1951, cho2007}, radiation \citep{tazaki2017}, or gas flow \citep{gold1952, lazarian2007, kataoka2019}. Since the grain alignment has been mainly studied in the condition of the interstellar medium, further discussion of the alignment physics in the protoplanteary disks is necessary.
Another mechanism that would produce polarization is the self-scattering, where the dust grains scatter and polarize the incoming thermal emission from other dust grains \citep{kataoka2015}. Dust grains with the size comparable to observed wavelengths efficiently scatter and polarize the thermal emission. The interpretations of the observations depend on each target. However, the interpretation itself is one of the major issues of the millimeter polarization of protoplanetary disks.

To interpret the dust polarization, it is essential to disentangle the polarization mechanisms. 
In this paper, as one step toward complete understanding of the polarization, we investigate the effects of superposition of the two major polarization mechanisms, which are the self-scattering and alignment. Especially, we model the 3.1 mm polarization of the HL Tau disk, which has been interpreted purely as alignment-induced polarization \citep{kataoka2017}, but also pointed out to be contaminated by self-scattering \citep{yang2019}.

HL Tau is a class I/I\hspace{-.1em}I protostar in the Taurus star-forming complex at a distance of 140 pc from the solar system \citep{rebull2004, robitaille2007}. A notable feature of the surrounding disk is the concentric multi-ringed structure with a spatial scale of $\sim$100 au \citep{alma2015}.
The polarization morphology at millimeter wavelengths on the HL Tau disk shows strong wavelength dependence \citep{stephens2014,kataoka2017, stephens2017}. 
The polarization morphology at 870 $\micron$ is parallel to the disk minor axis, while that at 3.1 mm shows almost the azimuthal pattern.
The polarization pattern with the vectors parallel to the minor axis is well reproduced by the self-scattering model \citep{yang2016, kataoka2016}.
Therefore, the $870~\micron$ polarization has been interpreted by the self-scattering.
In contrast, the origin of the 3.1 mm polarization has been interpreted by dust alignment because the self-scattering fails to explain the azimuthal pattern. 
It has been first interpreted by grain alignment with the radiative anisotropy \citep{tazaki2017,kataoka2017}, but slight inconsistency in the morphology has also been pointed out \citep{yang2019}. 

The switching of the polarization mechanisms between 870 $\micron$  and 3.1 mm must be explained. The reason why the self-scattering disappear at 3.1 mm is an expected behavior because polarization fraction quickly drops if dust grains are larger than the wavelengths divided by $2\pi$ \citep{kataoka2015}. However, this requires the grain size to be $\sim$ 100 $\micron$, which is not consistent with the general SED analysis, which requests millimeter to centimeter dust grains \citep[e.g.,][]{beckwith1991, ricci2010}. \citealt{carlos2019} analyzed the millimeter SED of the HL Tau disk with the effects of scattering-induced intensity reduction \citep{liu2019, zhu2019} and obtained grain size to be millimeter, which is larger than $\sim$ 100 \micron. In contrast, the reason why alignment-induced polarization are observed at 3.1 mm but disappear at 870 $\micron$ is still unresolved. One idea to explain the transition is that the grain size is comparable to the wavelengths, and their intrinsic polarization is strong enough to be detected at 3.1 mm but too weak at 870 $\micron$. This happens because of the transition from Rayleigh to Mie regime. If the grain size is much smaller than the wavelengths, the intrinsic polarization does not have strong wavelength dependence. If the grain size is much larger than the wavelengths, in contrast, the polarization fraction drops. Note that if the grain size is comparable to the wavelengths, polarization direction becomes perpendicular to the grain major axis because of the Mie regime \citep[e.g.,][]{guillet2020}. In this way, the wavelength dependence between millimeter and submillimeter wavelengths are the key to understanding the polarization.

Toward the full understanding of the wavelength-dependent polarization of the HL Tau,  in this paper, we focus on the modeling the 3.1 mm polarization by the mixture of the self-scattering and grain alignment.
\citet{yang2019} raised the issue that neither of aerodynamic nor radiative flux induced polarization can explain the 3.1 mm polarization and speculated that possible contamination of the self-scattering component. We quantitatively investigate how strong the self-scattering contamination is even at the 3.1 mm band and leave the full modeling of the three wavebands for future studies. 
Note that this does not solve the issue that no alignment model can perfectly explain the polarization, but it may change the wavelength dependence of the self-scattering components in the HL Tau polarization, which may have an impact on the grain size constraints.

\begin{figure*}[t]
\centering
\includegraphics[height=9cm]{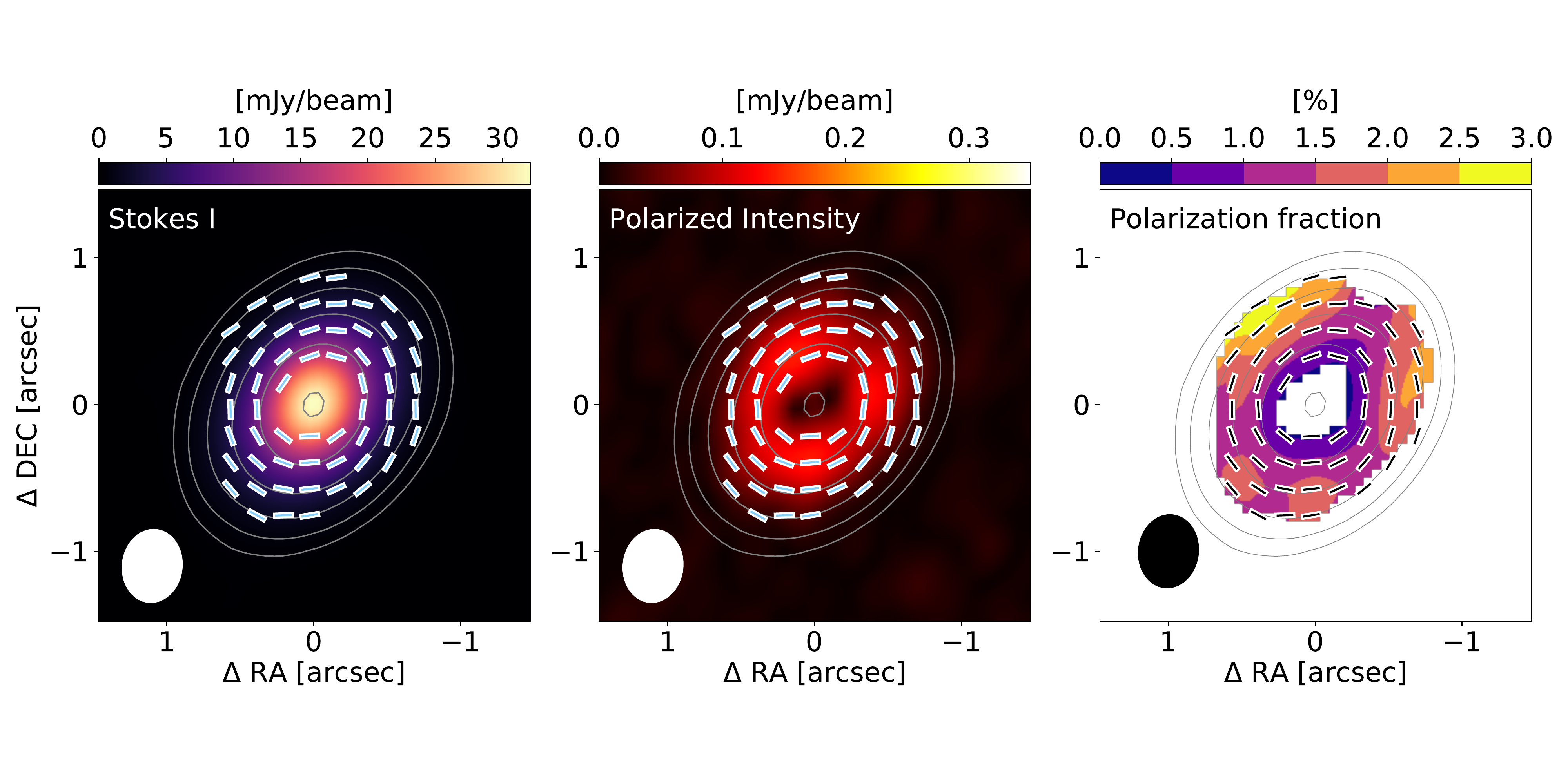}
\caption{ALMA Band 3 ($\lambda$ = 3.1 mm) polarimetric observations of the HL Tau disk, which has been previously reported by \citealt{kataoka2017} and \citealt{stephens2017}. The left panel shows the Stokes I intensity in color, the central panel shows the polarized intensity, and the right panel shows the polarization fraction.
The solid contours in the three panels are equivalent and represent the intensity with the levels of (10, 25, 60, 150, 366, 900) times the noise level, which is $\sigma_I$ (= 34 $\mu$Jy beam$^{-1}$). 
The overlaid line segments represent the polarization vectors where the polarized intensity is larger than 3 times the noise level of the polarized intensity, which is $\sigma_{PI}$ (= 5.3 $\mu$Jy beam$^{-1}$).
Note that the length of the polarization vectors is set to be equal. 
The beam has a size of 0.\arcsec51 $\times$ 0.\arcsec41 and position angle of -6\fdg76, which is represented in the bottom left as an ellipse. 
}
\label{fig:HLTau_Band3}
\end{figure*}

\section{Model} \label{sec:method}
We briefly describe the outline of the radiative transfer models in this study. First, we construct a dust disk model that reproduces the Stokes I emission obtained with the 3.1 mm polarization observation. Then, on the constructed disk model, we take into account the effect of the grain alignment and self-scattering in calculating polarization.

\subsection{Disk Model}
Figure \ref{fig:HLTau_Band3} shows the Stokes I intensity, polarized intensity, and polarization fraction, all of which are overlaid with the polarization vectors. The notable features of the 3.1 mm polarization are (1) azimuthal polarization morphology and (2) azimuthally uniform polarization fraction at $\sim$ 1\%. 

We construct an axisymmetric disk model that reproduces the stokes I image obtained on the polarization observation at 3.1 mm \citep{kataoka2017, stephens2017}. 
By assuming that the Stokes I emission at 3.1 mm is optically thin, we construct the radial profile of the optical depth at 3.1 mm with the form of
\begin{equation}
\rm \tau_{model} (r) = \tau_0 \: \Bigl(\frac{r}{\rm 1 \: au}\Bigr)^{-\gamma_1}\exp\Bigl[\Bigl(\frac{r}{r_c}\Bigr)^{-\gamma_2}\Bigr],
\label{eqn:tau_profile_model}
\end{equation}
with the temperature profile of $\rm T(r) = 310 (r/1 AU)^{-0.57} $ K \citep{okuzumi2019}. 
We note that both the dust surface density and temperature are assumed to have smooth radial distributions without any ring and gap structures observed with the higher spatial resolution \citep{alma2015} for simplicity. 



\begin{figure*}
\centering
\includegraphics[height=9cm]{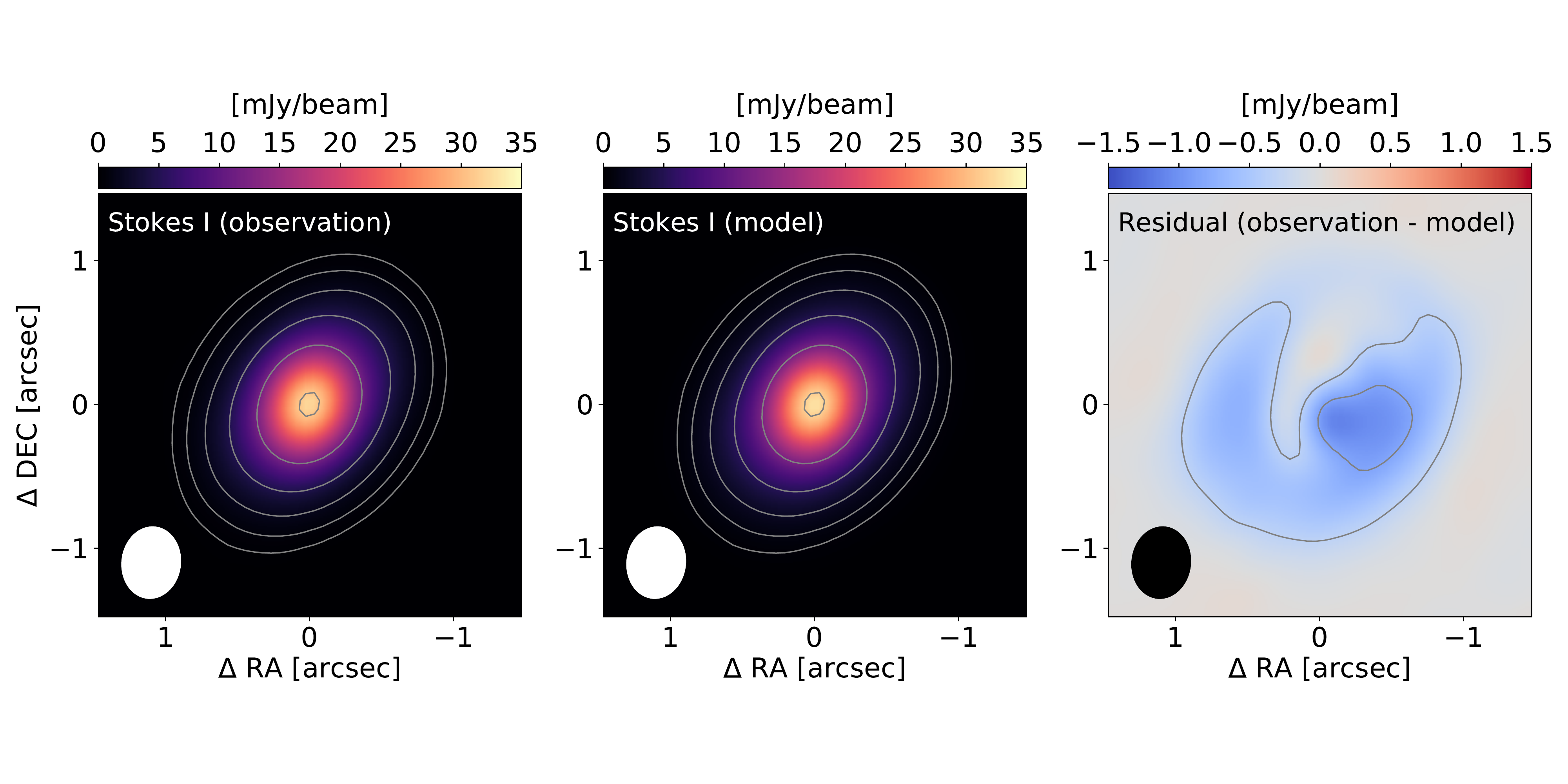}
\caption{The left panel shows the Stokes $I$ image of the observations, the central panel shows the model image, and the right panel shows the residual map where the observed Stokes $I$ subtracted with the model image. 
The contours of the left and the central panels are the same as Fig.\ref{fig:HLTau_Band3} and the contour of the right panels show the residual levels of (-10, -25) $\times$ $\sigma_I$}.
\label{fig:HLTau_observation_model_stokesI}
\end{figure*}

We adopt the parameter set of ($\rm \tau_0, r_c, \gamma_1, \gamma_2$) = (1.0, 129.2, 6.6, 9.0), which reproduces the Stokes I emission.
Figure \ref{fig:HLTau_observation_model_stokesI} shows the images of observed and modeled Stokes I produced assuming the parameter set and residual.

We derive the dust surface density profile $\rm \Sigma_d (r)$ by calculating $\rm \Sigma_d (r)$ = $\rm \tau_{model}/\kappa_{abs}$, where $\rm \kappa_{abs}$ is the absorption opacity.
As we will describe below, the value of $\kappa_{abs}$ is a function of $a_{max}$ and $\lambda$ when constituent materials are determined.
For the fiducial value, we adopt $\rm \kappa_{abs}(a_{max} = 100 \: \micron, \lambda = 3.1 \: mm)$ = 0.0788 cm$^2$ g$^{-1}$ in calculating the dust surface density.

The vertical dust density is assumed to be Gaussian density distribution with a dust scale height $h_d$ such that $\rm \rho_d = \Sigma_d/\sqrt{2\pi}h_d \exp(-z^2/2h_d^2)$. We also assume that $h_d$ is same as the gas pressure scale height.

While the residuals are still significant, they show axisymmetric distribution. For this study, we aim at reproducing the continuum emission with an axisymmetric model for simplicity, and thus we do not further search the parameter sets.

\subsection{Dust Model} 
We use a grain composition model developed by \citet{birnstiel2018}, where the grains are the mixture of silicate, troilite, organics, and water ice. The refractive index used in the calculation is as follows: \citet{draine2003} for astronomical silicate, \citet{henning1996} for troilite and refractory organics, and \citet{warren2008} for water ice. We compute the mixture of them with the effective medium theory using the Maxwell-Garnett rule (e.g., \citet{bohren1983, miyake1993}).
We assume the grains have a power-law-size distribution with a power of $q$ = -3.5 \citep{mathis1977} with the maximum grain size of $a_{max}$, while the minimum grain size is fixed at $a_{min}$ = 0.05 $\mu$m. 
We assume the same grain population throughout the disk.

To reproduce polarization, we consider both alignment and scattering-induced polarization.
For the alignment-induced polarization, we consider both prolate and oblate grains.
\footnote{Grains are believed to be spinning when they are aligned. The spinning motion makes the grains to be effectively prolate or oblate ellipsoidal bodies. We call prolate/oblate grains for simplicity but this does not rule out any complex shapes of the dust grains.}
The motivation to consider the both models is the observed polarization morphology. There are two ways to reproduce azimuthal polarization vectors; effectively prolate grains with their major axis to be parallel to the disk azimuthal direction and effectively oblate grains with their minor axis parallel to the disk radial direction. 
We assume the Rayleigh regime \citep{lee1985} to calculate the intrinsic polarization.
This approximation holds when the size parameter $x=2\pi a/  \lambda \ll 1$. 
We assume the grain size up to 200 $\micron$ at the wavelength of 3.1 mm, which corresponds to the size parameter of $x=0.42$, which justifies the use of Rayleigh regime. 
We assume the axes ratio of the grains to be $\alpha = 0.9$ for prolate grains and $\alpha = 1.1$ for oblate grains and that the alignment efficiency to be $\epsilon=0.6$. 
We assume that the oblate grains are aligned with their short axis parallel to the radial direction of the disk while the prolate grains are aligned with their long axis parallel to the azimuthal direction of the disk.
We ignore the effects of temperature for simplicity.

As pointed out by \citet{guillet2020}, to discuss the wavelength dependence of the alignment-induced polarization, the Mie regime is essential to be considered. In this paper, however, we do not treat the Mie regime for simplicity and we focus on the single band observations on the HL Tau at 3.1 mm. In the cases of further modeling at 870 $\micron$ and at 1.3 mm, the Mie regime treatment is required because the size parameter of $200~\micron$ grains becomes $x=1.44$ at 870 $\micron$ and $x=0.97$ at 1.3 mm.

The scattering-induced polarization is included by turning on the scattering in the radiative transfer simulations. We treat the full phase function and take into account the multiple scattering.
The controlling parameter of the scattering is the maximum grain size, $a_{max}$.
The self-scattering polarization is the most effective when the maximum grain size is $a_{max}$ $\sim$ $\lambda/2\pi$ \citep{kataoka2015}.

We assume spherical dust grains for computing self-scattering while we assume oblate or prolate grains for computing alignment-induced polarization.
These are not consistent but we neglect the effects of scattering by non-spherical grains for simplicity.
The angle-averaged absorption opacity of oblate/prolate grains is set to be the same as the spherical grains.

We also note a technical treatment. The radiative transfer code RADMC-3D allows users to compute the alignment-induced polarization only when the dust scattering is turned on. To reproduce the alignment-induced polarization without scattering, therefore, we use $a_{max}$ = 50 $\mu$m with the scattering turned on, where the scattering-induced polarization is negligible.

\section{Results} \label{sec:results}

\subsection{Alignment-only Model}
\begin{figure*}
\centering
\includegraphics[height=12cm]{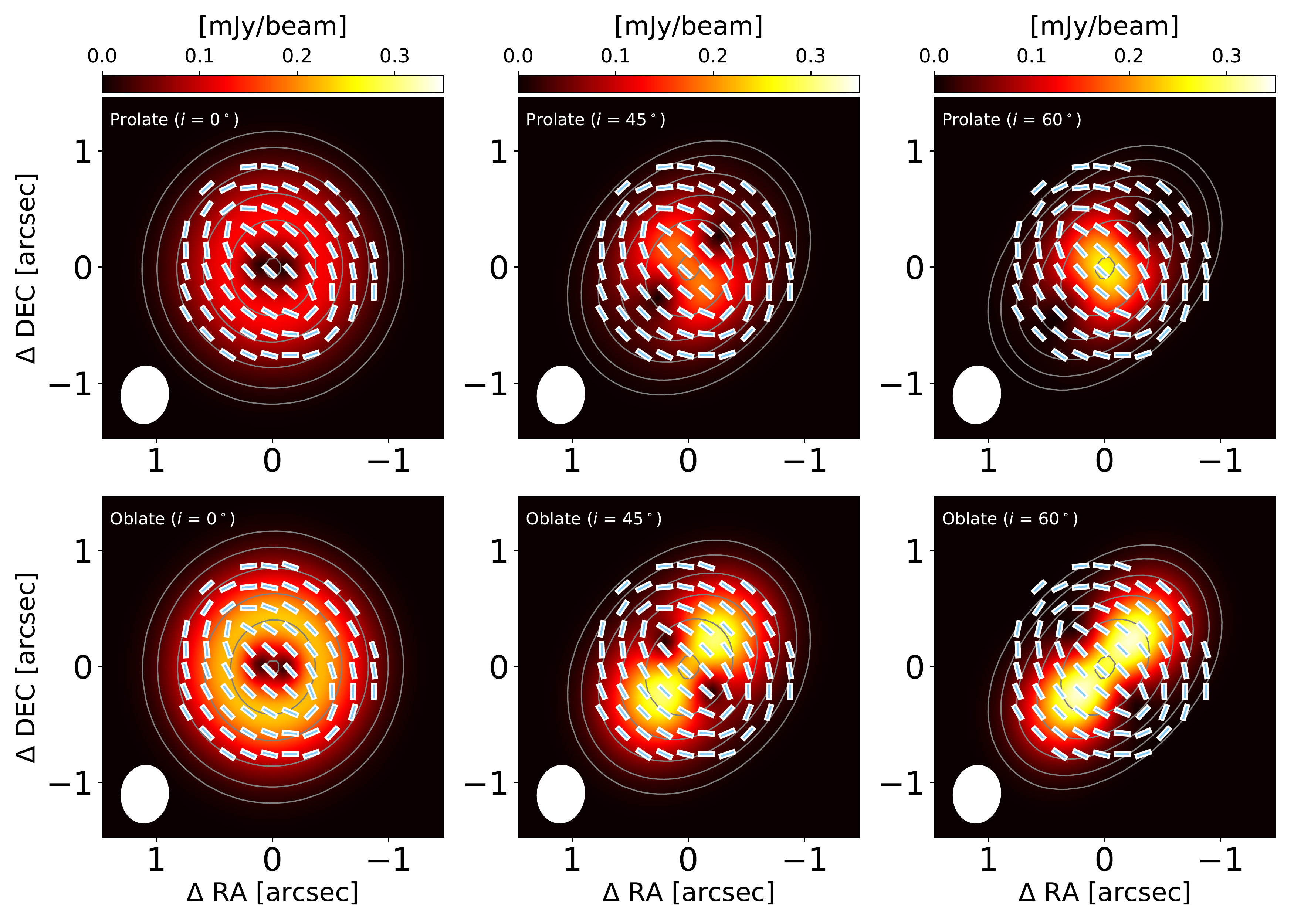}
\caption{The six panels show the polarized intensity of models with prolate and oblate grains with different disk inclinations.
The top three panels show the results of the prolate grains with their long axis parallel to the azimuthal direction and the disk inclinations are $i$ = 0$^\circ$ (left), 45$^\circ$ (center), and 60$^\circ$ (right).
The bottom three panels show the same as the top but for oblate grains.
The levels of contours and the line segments are the same as Fig.\ref{fig:HLTau_Band3} but for each model.
The synthesized beam is represented as the ellipse at the bottom left of each panel.}
\label{fig:HLTau_inclination_PI}
\end{figure*}

First, we see how the beam dilution affects the appearance of the polarization vectors. We use the same disk model of HL Tau but assume that the disk is observed with a face-on view, $i=0^\circ$. The left panels of Figure \ref{fig:HLTau_inclination_PI} demonstrates the effects of beam dilution with the face-on view. To reproduce the azimuthal polarization pattern, the prolate grains are assumed to be aligned with their long axis to be parallel to the azimuthal direction, while the oblate grains are with their short axis to be parallel to the radial direction. Both models generally shows the azimuthal pattern in polarization vectors, but the vectors at around the central region have certain inclination angles, which is not expected in the case without beam dilution.

Next, we see the effects of changing the disk inclination angle on the appearance of the polarization vectors and polarized intensity. The center and right columns of Figure \ref{fig:HLTau_inclination_PI} show the case of the inclination angles of 45$^\circ$, and 60$^\circ$.
The polarization vectors are almost in the azimuthal directions in the inclination angles of 45$^\circ$ and 60$^\circ$ for both the prolate and oblate grains, except the central regions where the beam dilution affect the morphology. 

\citet{yang2019} proposed that, if the disk inclination is non-zero, we can distinguish the alignment models between effectively prolate and oblate grains by investigating if the polarization vectors are in the azimuthal directions (i.e., azimuthal pattern) or are with their normal direction to be toward the disk center (i.e., circular pattern). 
However, as we see in Figure \ref{fig:HLTau_inclination_PI}, the difference between elliptical and circular patterns is hard to be distinguished because of the beam dilution.

In contrast, the polarized intensity shows prominent differences between the prolate and oblate grain models; the prolate grain models show higher polarized intensity (i.e., brighter color in Figure \ref{fig:HLTau_inclination_PI}) along the minor axis than the major axis, while the oblate models show higher polarized intensity along the major axis.
The reason can be understood by a simple geometry. Prolate grains along the disk minor axis always shows the longest axis in the observed projected plane even if the disk inclination is changed, while those along the disk major axis shows less polarization fraction if the disk inclination changes to be closer to edge-on view because of the projection effects. The situation is completely opposite in the case of oblate grains. Oblate grains along the disk major axis always shows the longest axis to the observers while those along the minor axis becomes fainter in higher inclination angles because of the projection effects. Therefore, investigating the polarized intensity pattern is a promising way to distinguish the mechanisms between prolate and oblate models if the polarization is purely due to the grain alignment.

Now we come back to the case of the HL Tau disk. As shown in the central panel of Figure \ref{fig:HLTau_Band3}, the polarized intensity of the HL Tau disk at 3.1 mm wavelength shows little variation in the azimuthal direction. This is different from the models that we discussed, where the prolate grain model shows that polarized intensity is brighter along the disk minor axis while the oblate model shows it is brighter along the disk major axis. This illustrates that alignment-only models may not be able to reproduce the observed polarized intensity pattern. This motivates us to add another mechanism of polarization, self-scattering, which we will discuss in the next chapter.

We note that the general properties described above hold for the polarization fraction as shown in Figure \ref{fig:HLTau_main_PF}. In addition, the general behavior and differences between prolate and oblate grain models is also discussed by \citet{yang2019} with semi-analytical models.

\subsection{Mixture of Alignment and Scattering Models}
Next, we discuss a possibility that a mixture of the alignment and self-scattering models can reproduce the HL Tau polarization at 3.1 mm. 
The self-scattering polarization generally produces the polarization vectors parallel to the minor axis, and the fraction is a strong function of the grain size. Therefore, we set the maximum grain size $a_{max}$ to be the the major parameter and changes it to reflect the strength of the self-scattering polarization. We use the same disk models of the alignment-induced polarization of the prolate and oblate cases, and add the self-scattering-induced polarization.

\begin{figure*}
\centering
\includegraphics[height=12cm]{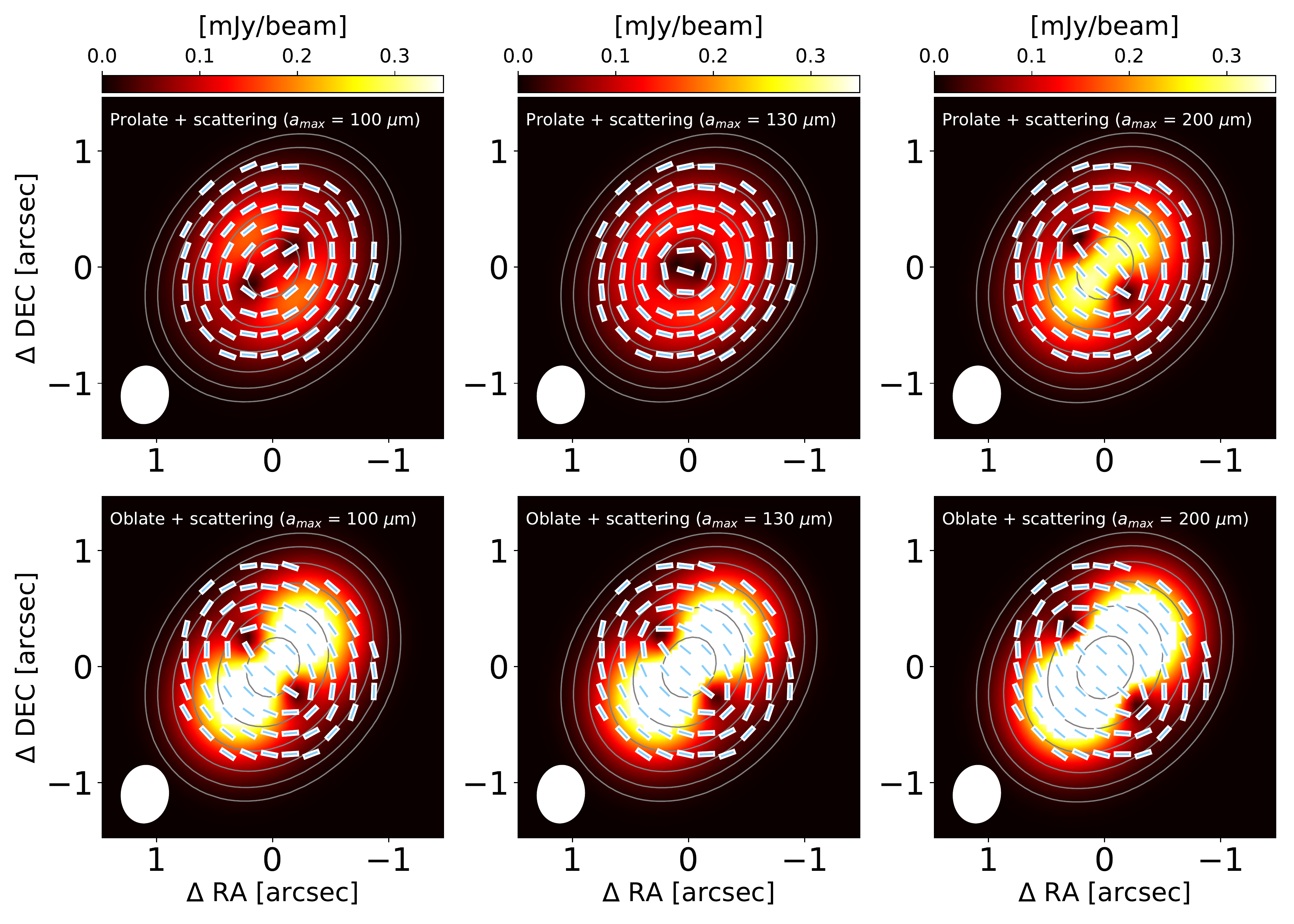}
\caption{
The six panels show the polarized intensity of models with mixture of the grain alignment and self-scattering. 
The upper three panels show the case of the combination of the aligned prolate grain and self-scattering.
The prolate grain model is fixed through the three panels while the self-scattering-induced polarization is changed by setting the maximum grain size to be $a_{max}$ = 100 $\micron$ (left), 130 $\micron$ (center), and 200 $\micron$ (right). 
Note that the contribution of self-scattering is weaker for the left panel and stronger for the right panel.
The bottom panels are the same as the top panels but for oblate grains.}
\label{fig:HLTau_main_PI}
\end{figure*}

Figure \ref{fig:HLTau_main_PI} shows the polarized intensity of the prolate and oblate grain models with $i$ = 45$^\circ$. The contributions of the self-scattering model increases by increasing $a_{max}$ from left to right panels while that of the alignment models are fixed.

As expected, the oblate + scattering models do not reproduce the observation. Both polarization of the alignment of oblate grains and self-scattering produces stronger polarization along the major axis. As a result of superposition, the azimuthal variation becomes stronger, which is not consistent with the observation as shown in Figure \ref{fig:HLTau_Band3}, which shows little azimuthal variation in polarized intensity.
Instead, the prolate + scattering model shows less azimuthal variation than the oblate + scattering model. This is because the prolate grain alignment produces strong polarization along the disk major axis while the self-scattering does along the disk minor axis. As a result of their superposition, the polarized intensity shows less azimuthal variation, which is consistent with the observations.

Now the question is how strong is the contributions of the alignment and scattering. We focus on the prolate + scattering model and discuss it. The top three panels of Figure \ref{fig:HLTau_main_PI} show the case of different grain sizes, which corresponds to difference of contribution of self-scattering. As demonstrated in the figure, the  $a_{max}$ = 100 $\mu$m case is dominated by the alignment-induced polarization because the polarized intensity is stronger along the minor axis, while the  $a_{max}$ = 200 $\mu$m is by the self-scattering because it is stronger along the major axis. The $a_{max}$ = 130 $\mu$m case shows the half and half contributions between alignment and scattering, which reproduce little azimuthal variation as observed. Therefore, we conclude that prolate + scattering model with $a_{max}$ = 130 $\mu$m best reproduces the observed HL Tau polarization at 3.1 mm.
Note that we also successfully reproduce the observed polarization fraction with the mixture model $a_{max}$ = 130 $\micron$ as shown in Figure \ref{fig:HLTau_main_PF}.

\section{Discussion} \label{sec:discussion}
The main results of the radiative transfer calculations are (1) the HL Tau 3.1 mm polarization can be reproduced with the mixture of the alignment and self-scattering models and (2) prolate grain model can explain the observations but oblate model does not.
In this section, we discuss the implications on the grain properties from both alignment and scattering theories as well as the caveats of our models.

\subsection{What can reproduce effectively prolate shape?} 

As discussed also by \citet{yang2019}, the effectively prolate grain model is more preferable for reproducing the HL Tau polarization but it is theoretically counter-intuitive. 
We usually assume that grains are spinning if they are aligned, and they are observed as effectively oblate grains.

One possible speculation for explaining the prolate grain model is that grains are not spinning in the HL Tau disk.
The only mechanism that does not assume grain spinning so far would be the mechanical alignment with a supersonic gas flow, which is so-called Gold alignment, where the long axis of needle-like grains is parallel to the gas flow onto dust grains \citep{gold1952}.
However, the Gold alignment assumes that the gas speed on the dust grains is supersonic whereas that is generally subsonic in protoplanetary disks \citep[e.g.,][]{cho2007}. 
All the discussion above is phenomenological and we need further development of general grain alignment theory in protoplanetary disks.

Furthermore, we note that, even if the Gold mechanism is at work, the predicted polarization pattern is not consistent with the observations.
The direction of the polarization would be parallel to the gas flow on the frame of dust grains \citep{kataoka2019}, which is a strong function of the Stokes number, which denotes how well dust grains couple to ambient gas.
The grain radius $a$ is found to be $a_{max}$ = 130 $\micron$ in our modeling, which gives the Stokes number to be St = 5 $\times$ 10$^{-3}$ with the assumption of the gas surface density of $\rm \Sigma_g$ $\sim$ 20 g cm$^{-2}$. 
With such a small number of St much smaller than unity, the directions of gas velocity on grains would be in the radial directions \citep[see Figure 2 in][]{kataoka2019}.
Note that this discussion holds even for ring-like disks \citep{mori2019}.
However, our modeling has found that the directions of the relative velocity are dominantly in the azimuthal directions, which is incompatible with the theory. 

\subsection{How Efficient is the Scattering at 3.1 mm?}

The size of dust grains has been constrained by modeling the wavelength dependence of the contribution of the scattering-induced polarization.
In the case of HL Tau disk, previous studies have assumed that the Band 7 (0.87 mm) polarization is dominated by self-scattering while the Band 3 (3.1 mm) has no contribution of self-scattering, which results in the tight constraint on the grain size to be 70 $\micron$ in size \citep{kataoka2016, yang2016, kataoka2017}.
However, we have revealed that there is a contribution of self-scattering even for the 3.1 mm polarization.
Therefore, we discuss if our modeling changes the interpretation of the grain size in the HL Tau disk.

To estimate the scattering component at each wavelength, as conducted in \citet{kataoka2017}, we compute the total polarization across the HL Tau disk, which is given by
\begin{eqnarray}
PF_{total} =  
\left\{ 
\begin{array}{ll}
1 & \rm{(minor)}\\
-1 &\rm{(major)}\\
\end{array}
\right\}
\times {\sqrt{(\Sigma Q)^2 + (\Sigma U)^2}}{\Sigma I}  , 
\label{eqn:total_polarization}
\end{eqnarray}
where $\Sigma$I, $\Sigma$Q, and $\Sigma$U are the Stokes I, Q, and U emissions integrated for the whole disk.
Here, we explicitly define the sign of the total polarization fraction; $PF_{total}$ is positive if the position angle of the total polarization, which is calculated from $\Sigma$Q, and $\Sigma$U, is closer to the disk minor axis, while $PF_{total}$ is negative if the position angle is closer to the disk major axis. This is motivated to be able to estimate the canceling out effects. In the case of axisymmetric protoplanetary disks, the position angle of polarization is usually parallel to either the disk minor or major axes because of the symmetry. If this is the case, defining sign with respect to the disk minor and major axes helps to see which components dominate the total polarization.

Table \ref{tab:scattering_contributions} summarizes the computed total polarization fraction for the observations and models.
We update the total polarization fraction presented in \citet{kataoka2017}, who used the SMA and CARMA data, with the ALMA data in Band 6 and 7 while its upper limit in ALMA 3 is presented the same as that study. $PF_{total}$ is calculated from the final images as shown in Figure \ref{fig:HLTau_main_PI} with the equation (\ref{eqn:total_polarization}). We also show $PF_{scat, model}$ and $PF_{align, model}$, which represent the total polarization of the cases where either scattering or alignment model is calculated. As shown in the table \ref{tab:scattering_contributions}, the total polarization fraction is nearly equal to the sum of those of the scattering and alignment components. This illustrates the powerful estimation of the total polarization by considering the positive and negative signs of each polarization components.

\begin{deluxetable*}{ccccc}[t]
\tablecaption{
Summary of the total polarization fraction of the models and observations. The first column represents the model names. $\lambda$ is the wavelengths. $PF_{total}$ represents the total polarization fraction, which can be decomposed to the scattering component, $PF_{scat}$ and the alignment component,  $PF_{align}$. We define positive values of the polarization fraction for the component parallel to the disk minor axis while negative values for that parallel to the disk major axis. \label{tab:scattering_contributions}
}
\tablecolumns{5}
\tablenum{1}
\tablewidth{0pt}
\tablehead{
\colhead{} & 
\colhead{$\lambda$ [mm]} & 
\colhead{$PF_{total} (\simeq PF_{scat}+PF_{align})$  [\%]} &
\colhead{$PF_{scat}$  [\%]} &
\colhead{$PF_{align}$ [\%]} 
}
\startdata
oblate                    & 3.1 & 1.07  & 0& 1.07 \\
oblate + scattering ($a_{max}$ = 100 $\micron$) & 3.1 &  1.25 &0.20 &  1.07\\
oblate + scattering ($a_{max}$ = 130 $\micron$)  & 3.1 & 1.47  & 0.42 & 1.07 \\
oblate + scattering ($a_{max}$ = 200 $\micron$)  & 3.1 &  2.23 & 1.21& 1.07 \\
prolate                  & 3.1 & -0.55  & 0& -0.55 \\
prolate + scattering ($a_{max}$ = 100 $\micron$)  & 3.1 &  -0.35 & 0.20&  -0.55\\
prolate + scattering ($a_{max}$ = 130 $\micron$)  & 3.1 &  -0.12 & 0.42 & -0.55 \\
prolate + scattering ($a_{max}$ = 200 $\micron$)  & 3.1 & 0.68  &1.21 &  -0.55\\
\hline
observation & 3.1   & $-0.1<PF_{total}<0.1$  & $\--$    & $\--$ \\ 
observation &1.3   & $0.551 \pm 0.003$       & $\--$    & $\--$ \\
observation &0.87 & $0.619 \pm 0.004$       & $\--$    & $\--$ \\ 
\enddata
\end{deluxetable*}

As shown in the table \ref{tab:scattering_contributions}, oblate grain model always shows the positive value while the prolate grain model shows the negative values, where positive value corresponds to the polarization parallel to the disk minor axis while negative value corresponds to that parallel to the disk major axis. In contrast, the self-scattering always shows positive values because of the disk geometry. To reproduce the tight upper limit on the absolute value of the total polarization fraction to be less than 0.1 \%, the oblate grain model is not favored and the combination of the prolate model and self-scattering is the key.

Here, we emphasize our big assumption that we do not know the wavelength dependency of the alignment-induced polarization, and assume that there is no contribution from alignment polarization at the wavelengths of the Band 6 and 7. To reproduce this, we need to calculate the intrinsic polarization in the Mie regime \citep{guillet2020,kirchschlager2019}, which computationally costs.

By assuming that there is no contribution of alignment induced polarization at Bands 6 and 7, we confirm the decreasing trend of the observed total polarization fraction from 0.87 mm to 3.1 mm, which has been discussed by \citet{kataoka2017} partially with the SMA and CARMA results from \citep{stephens2014}.
Furthermore, our study revealed that the non-detection of total polarization at 3.1 mm is explained by the combination of the alignment (-0.55 \%) and the self-scattering ($+0.42$ \%), which requires modification on the wavelength-dependence of the self-scattering polarization.
However, the contribution of self-scattering to the total polarization at 3.1 mm, which is estimated to be 0.42 \% in this study, is still smaller than the 0.87 mm and 1.3 mm total polarization fractions.
This means that the decreasing trend of the self-scattering polarization still holds, which supports the idea of $\sim 100$ micron-size grains because the wavelength, at which the self-scattering polarization peaks is, 0.87 mm or shorter \citep[see][]{kataoka2016, kataoka2017, yang2016}.

Still, there are significant differences in the size measurements between millimeter-wave polarization and spectral indices.
Latest modeling for the continuum emission of the HL Tau disk obtained at multiple wavelengths from 0.9 mm to 13 mm found the radial gradient of the grain sizes from $a_{max}$ = 1.5 mm in the inner region to $a_{max}$ = 500 $\micron$ in the outer region, which is at least four times larger than that in this study \citep{carlos2019}. 
Plenty of differences between the two studies, for example, the observation wavelengths, angular resolutions, and modeling methods can affect the size measurements and partly explain the discrepancy.
Another primary difference is assumed constituent materials of the dust grains, which determine refractive indices. Because the absolute values of the absorption/scattering opacities and their wavelength dependence are strongly sensitive to the assumption on the composition, this can affect the measurement conducted in both of the studies \citep{testi2014}.
To reconcile the discrepancy, further modeling that assumes the same grain composition and disk parameters should be conducted.

\subsection{Caveats}

In the radiative transfer calculations, the dust models are not fully consistent between the alignment and scattering models.
When we calculate the polarization of aligned elongated grains, we set an angle-dependent absorption opacities.
However, we assume spherical dust grains that have the same absorption opacities for elongated grains for calculating their scattering because of the difficulties in calculating scattering opacities of non-spherical dust grains.

\section{Conclusions} \label{sec:conclusions}
The origin of the polarized emission of the HL Tau disk at 3.1 mm has been debated in the previous works. 
We have focused on the 3.1 mm polarization of HL Tau, which shows the azimuthal pattern of polarization vectors, and have conducted radiative transfer simulations for the 3.1 mm HL Tau polarization. Our main findings are as follows.
\begin{enumerate}
\item By radiative transfer simulations, we confirmed that alignment-induced polarization cannot reproduce the 3.1 mm polarization feature of the HL Tau disk as discussed by \citet{yang2019} with a semi-analytical model.
\item Next, we run models of the mixture of the grain alignment and self-scattering and found that the combination of the effectively prolate grain emission and self-scattering can reproduce the observations. However, the combination of the effectively oblate grain emission and self-scattering cannot reproduce the polarization. 
Our model shows that the total polarization fraction at 3.1 mm can be understood by the combination of 0.42 \% polarization due to the self-scattering, which is parallel to the disk minor axis, and 0.55 \% of the alignment-induced polarization, which is parallel to the disk major axis.
\item We found that the maximum size of the scattering dust grains is $a_{max}$ $\sim$ 130 $\micron$ to reproduce the azimuthally uniform polarization fraction. This size is consistent with them obtained on the polarization observation in Band 6 and 7. However, the size is a few times smaller than that was measured by the multi-wavelength continuum analysis \citep{carlos2019}. 
We confirmed that effectively prolate grain models are more preferable than the oblate grain models to reproduce the 3.1 mm polarization of HL Tau. However, the physical model to reproduce the situation is still uncertain.
\end{enumerate}

We note that while the radiative transfer calculations have been performed for explaining the 3.1 mm polarization of the HL Tau disk, further modeling of the multi-wavelength polarization with the results at 870 $\micron$ and 1.3 mm is required in future studies.
Modeling of the multiband polarization observations will provide the spectrum of alignment and scattering efficiency, which enable us to further constrain the grain sizes and shapes.

\acknowledgments
{We gratefully thank Ian Stephens for providing us the analyzed ALMA Band 3 polarization data, which has enabled us to conduct the modeling work.
We are also deeply grateful to Takashi Miyata, Takafumi Kamizuka, and Ryou Ohsawa for fruitful discussions for our study.
This work is supported by JSPS KAKENHI Numbers 18K13590 and 19H05088.
ALMA is a partnership of ESO (representing its member states), NSF (USA) and NINS (Japan), together with NRC (Canada), MOST and ASIAA (Taiwan), and KASI (Republic of Korea), in cooperation with the Republic of Chile. The Joint ALMA Observatory is operated by ESO, AUI/NRAO, and NAOJ.
}


%

\vspace{5mm}
\facilities{ALMA}





\appendix
\section{The Polarization Fraction in the 
Mixture Models (Alignment + Self-scattering)} \label{sec:mixture_PF}
\begin{figure*}
\centering
\includegraphics[height=12cm]{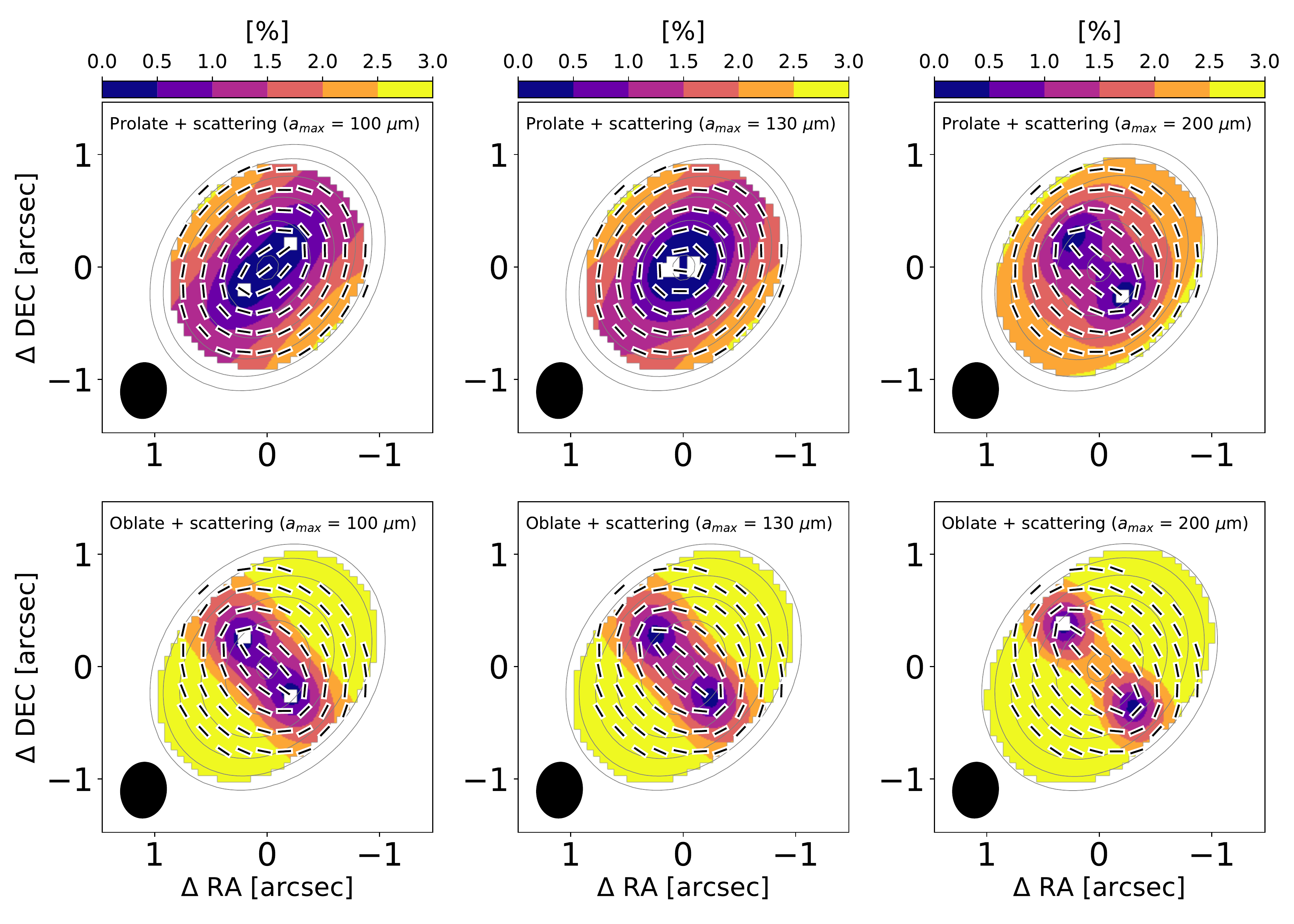}
\caption{Same as Figure \ref{fig:HLTau_main_PI} but for polarization fraction.}
\label{fig:HLTau_main_PF}
\end{figure*}

Figure \ref{fig:HLTau_main_PF} shows polarization fraction mixture of the grain alignment and self-scattering. As well as Figure \ref{fig:HLTau_main_PI} in Section \ref{sec:results}, the model with prolate grains and $a_{max}$ = 130 $\micron$ (top-central panel) successfully reproduces the azimuthally uniform polarization fraction with $\sim$1$\--$2 \%, whereas other models predict significant azimuthal variations.

\bibliography{sample63}{}
\bibliographystyle{aasjournal}



\end{document}